# Generation of Multipartite Continuous-variable Entanglements via Atomic Spin Wave

Xihua Yang[1], Yuanyuan Zhou[1], and Min Xiao[2,3]

[1]*Department of Physics, Shanghai University, Shanghai 200444, China*

[2]*National Laboratory of Solid State Microstructures and Department of Physics, Nanjing University, Nanjing 210093, China*

[3]*Department of Physics, University of Arkansas, Fayetteville, Arkansas 72701, USA*

We present a proof-of-principle way to generate nondegenerate multipartite continuous-variable entanglements via atomic spin wave induced by the strong coupling and probe fields in the Λ-type electromagnetically induced transparency configuration in an atomic ensemble. Quantum correlated/anti-correlated and entangled Stokes and anti-Stokes fields, simultaneously produced through scattering the applied laser fields off the atomic spin wave, can be achieved. This method can, in principle, be extended to flexibly and conveniently create multicolor multipartite entangled narrow-band fields to any desired order with long correlation time, which may find promising applications in quantum information processing and quantum networks.

PACS numbers: 42.50.Gy, 03.67.Bg, 42.50.Dv, 42.65.Lm

Generating multipartite continuous-wave (CV) entanglements is a key resource for the implementation of quantum information protocols [1-3]. The most commonly-used tool is to employ the linear optical elements, i.e., polarizing beam splitters, to combine squeezed beams to produce multipartite quantum correlations and entanglements [4-5]. However, such produced entangled multiple fields are degenerate and suffer from short correlation time, thus limiting their potential applications in quantum memory and quantum networks [3]. Recently, generations of multipartite CV entangled fields with different frequencies have been examined by using nonlinear optical processes. Nussenzveig *et al.* demonstrated the creation of three-color (pump, signal, and idler) entanglements in the above-threshold optical parametric oscillator [6]; other ways have been proposed by applying either cascaded nonlinearities or concurrent parametric oscillation [7, 8]. Yet, these schemes are still suffering from having relatively short correlation time. An alternative promising avenue is the implementation of nonlinear processes in atomic media to mediate the creation of multi-entangled fields. By using nondegenerate four-wave mixing (FWM) or Raman scattering processes in an atomic ensemble, the electromagnetically induced transparency [9] (EIT)-based double-$\Lambda$-type systems have been actively studied for efficiently creating nondegenerate quantum correlated/entangled narrow-band photon pairs [10-12], where the correlation time is determined by the long coherence decay time ($\sim ms$ or even $\sim s$) between the two lower states, thereby having the virtue suitable for quantum memory required in quantum communication [10]. However, so far no scheme has been proposed to create arbitrary number of nondegenerate CV entangled fields in an atomic ensemble.

In this Letter, we propose a convenient and flexible way to create multicolor multipartite CV entanglements via an atomic spin wave established by the strong on-resonant coupling and probe fields in the $\Lambda$-type EIT configuration, motivated by our experimental observation of generating classical multi-field correlations and anti-correlations via atomic spin coherence in $^{85}$Rb atomic system [13]. Multiple entangled Stokes and anti-Stokes fields, simultaneously produced through scattering the applied laser fields off the atomic spin wave, can be obtained by using an atomic ensemble. This method can, in principle, provide an alternative way to create nondegenerate multi-entangled CV fields (up to arbitrarily high order) with long correlation time, which may find interesting applications in quantum communication and quantum information processing.

The considered model, as shown in Fig. 1a, is based on our experimental configuration used in Ref. [13], where the relevant energy levels and the applied/generated laser fields form a quintuple-$\Lambda$-type system. Levels $|1\rangle$, $|2\rangle$, and $|3\rangle$ correspond, respectively, to the ground-state hyperfine levels $5S_{1/2}$ (F=2), $5S_{1/2}$ (F=3), and the excited state $5P_{1/2}$ in $D_1$ line of $^{85}$Rb atom with the ground-state hyperfine splitting of 3.036 GHz. The probe field $E_p$ (with frequency $\omega_p$ and Rabi frequency $\Omega_p$) and coupling field $E_c$ (with frequency $\omega_c$ and Rabi frequency $\Omega_c$) are relatively strong and tuned to resonance with the transitions $|1\rangle$-$|3\rangle$ and $|2\rangle$-$|3\rangle$, respectively. A third mixing field $E_m$, which can be generated from the coupling field with an acousto-optic modulator, off-resonantly couples levels $|2\rangle$ (or $|1\rangle$) and $|3\rangle$. As previously shown [13, 14], two Stokes fields $E_1$ & $E_3$ (with frequencies $\omega_1$ & $\omega_3$) and two anti-Stokes fields $E_2$ & $E_4$ (with frequencies $\omega_2$ & $\omega_4$) can be simultaneously created through nondegenerate FWM processes with the coupling, probe, and mixing fields acting on both $|1\rangle$-$|3\rangle$ and $|2\rangle$-$|3\rangle$ transitions at high atomic density and high laser powers; that is, two coupling (probe) photons are converted into one Stokes $E_1$ (anti-Stokes $E_2$) photon and one probe (coupling) photon; also one probe (coupling) photon and one mixing photon are absorbed and one coupling (probe) photon and one anti-Stokes $E_4$ (Stokes $E_3$) photon are emitted. In fact, the above simultaneously generated four FWM fields can be equivalently viewed as scattering the coupling, probe, and mixing fields off the atomic spin wave (S) pre-established by the strong (on-resonant) coupling and probe fields in the $\Lambda$-type EIT configuration formed by levels $|1\rangle$, $|2\rangle$, and $|3\rangle$, where the induced spin wave acts as a frequency converter with frequency equal to the separation between the two lower states [15]. The equivalent configuration is shown in Fig. 1b, which can be readily generalized to the N-Λ-type (with N being a positive integer) system by applying more laser fields tuned to the vicinity of the transitions $|1\rangle$-$|3\rangle$ and/or $|2\rangle$-$|3\rangle$ to mix with the induced atomic spin wave.

We first investigate the generation of quantum anti-correlated/entangled Stokes field $E_1$ and anti-Stokes field $E_2$ via atomic spin wave in a triple-Λ-type system by blocking the mixing field $E_m$. We employ the equivalent configuration in Fig. 1b to treat the generated Stokes and anti-Stokes fields. We assume that the Rabi frequencies of the scattering fields (i.e., the off-resonant coupling and probe fields) are far smaller than their frequency detunings, so the coupling between different scattering fields can be neglected and the Heisenberg equations can have linear dependence on the scattering fields; also, the generated Stokes and anti-Stokes fields are assumed to be very weak as compared to the scattering fields, thus, the scattering fields can be treated classically, whereas the Stokes field $E_1$, anti-Stokes field $E_2$, and atomic spin field S can be treated quantum mechanically. After adiabatic elimination of the upper excited state, the effective Hamiltonian of the system in the interaction picture has the form [15,16]

$$H_I = \hbar \left[ k_1 \left( a_1^+ S^+ + a_1 S \right) + k_2 \left( a_2^+ S + a_2 S^+ \right) \right],$$

where $S = (1/\sqrt{N_a}) \sum_i |1\rangle_i \langle 2|$ is the collective atomic spin field with $N_a$ as the total number of atoms in the interaction volume. $k_{1,2} = g_{23,13} \Omega_{p,c} \sqrt{N_a} / \Delta_{1,2}$ with $\Delta_1 = \omega_1 - \omega_{32} = \omega_c - \omega_{31}$ ($\Delta_2 = \omega_2 - \omega_{31} = \omega_p - \omega_{32}$) as the detuning of the Stokes (anti-Stokes) field from the resonant transition $|2\rangle$-$|3\rangle$ ($|1\rangle$-$|3\rangle$). $g_{23}$ ($g_{13}$) is the coupling coefficient between the Stokes (anti-Stokes) field and its respective atomic states. Following Ref. [16], we define the exchange constant of motion ($C_1$) as

$$C_1 = k_1 \left( a_1^+ S^+ + a_1 S \right) + k_2 \left( a_2^+ S + a_2 S^+ \right).$$

By solving the Heisenberg equations of motion for the operators, the equation of motion for the atomic spin operator S can be expressed as

$$\frac{d^2 S}{dt^2} - 2iC_1 \frac{dS}{dt} - (k_1^2 - k_2^2) S = 0.$$

We set $\beta = \sqrt{C_1^2 - (k_1^2 - k_2^2)}$ and obtain the solutions for the operators as functions of their initial values for the case of $k_1 \neq k_2$:

$$S(t) = e^{iC_1 t} \left[ \left( \cos(\beta t) + \frac{iC_1}{\beta} \sin(\beta t) \right) S(0) - \frac{ik_1}{\beta} \sin(\beta t) a_1^+(0) - \frac{ik_2}{\beta} \sin(\beta t) a_2(0) \right],$$

$$a_1(t) = -\frac{ik_1}{\beta} \sin(\beta t) e^{-iC_1 t} S^+(0) + \frac{-k_1^2 \beta + k_1^2 \beta \cos(\beta t) e^{-iC_1 t} + ik_1^2 C_1 \sin(\beta t) e^{-iC_1 t} + \beta(C_1^2 - \beta^2)}{\beta(C_1^2 - \beta^2)} a_1(0)$$

$$+ \frac{-k_1 k_2 \beta + k_1 k_2 \beta \cos(\beta t) e^{-iC_1 t} + ik_1 k_2 C_1 \sin(\beta t) e^{-iC_1 t}}{\beta(C_1^2 - \beta^2)} a_2^+(0),$$

$$a_2(t) = -\frac{ik_2}{\beta} \sin(\beta t) e^{iC_1 t} S(0) + \frac{k_1 k_2 \beta - k_1 k_2 \beta \cos(\beta t) e^{iC_1 t} + ik_1 k_2 C_1 \sin(\beta t) e^{iC_1 t}}{\beta(C_1^2 - \beta^2)} a_1^+(0)$$

$$+ \frac{k_2^2 \beta - k_2^2 \beta \cos(\beta t) e^{iC_1 t} + ik_2^2 C_1 \sin(\beta t) e^{iC_1 t} + \beta(C_1^2 - \beta^2)}{\beta(C_1^2 - \beta^2)} a_2(0).$$

For the case of $k_2 = k_1$, as analyzed in Ref. [7], the solutions for the operators can be obtained by using the stochastic integration, which will be discussed in a subsequent paper. As seen in Fig. 1b, the collective atomic state is initially in a coherent superposition state, and the Stokes and anti-Stokes fields are initially in vacuum, so the initial state of the atom-field system can be written as $|\varphi_0\rangle = \sum_i (\frac{1}{\sqrt{2}} |1, 0_s, 0_{as}\rangle_i + \frac{1}{\sqrt{2}} |2, 0_s, 0_{as}\rangle_i) / \sqrt{N_a}$ (here we assume $\Omega_p = \Omega_c$ for simplicity), where $|1, 0_s, 0_{as}\rangle$ ($|2, 0_s, 0_{as}\rangle$) represents an atom in state $|1\rangle$ ($|2\rangle$) and the Stokes and anti-Stokes fields in vacuum. We use the criterion $V = (\Delta u)^2 + (\Delta v)^2 < 4$ proposed in Ref. [17] to verify the two-field entanglement of the generated Stokes and anti-Stokes fields, where $u = x_1 + x_2$ and $v = p_1 - p_2$ with $x_j = (a_j + a_j^+)$ and $p_j = -i(a_j - a_j^+)$.

Figures 2a-2c show the interaction time evolutions of V under different $k_2$ values with $k_1 = 1$ and $C_1 = 30k_1$. It can be seen that, under different $k_2$ values, V, with the initial value of 4, evolves with interaction time and becomes less than 4, which is a sufficient indication that genuine bipartite entanglement is produced; in addition, V exhibits an oscillation as a function of interaction time with the period of $T = \frac{2\pi}{C_1 - \beta} \doteq \frac{4\pi C_1}{|k_1^2 - k_2^2|}$ (for $C_1 \gg k_1, k_2$). It is interesting to note that, as shown in Fig. 2b (with $k_2$ about equal to $k_1$), there exists certain interaction time (at about t=T/2) that the minimum value of V nearly equals zero, which indicates the creation of two perfectly squeezed fields. The increase or decrease of $k_2$ would lead to the increase of the minimal value of V (see Figs. 2a and 2c), i.e., the degree of bipartite entanglement would be weakened. Further calculations show that the average photon numbers of the generated Stokes and anti-Stokes fields also oscillate with the same period as V, and the maximal bipartite entanglement takes place when the average photon numbers of the Stokes and anti-Stokes fields reach their peak values.

The generated bipartite quantum anti-correlations and entanglements can be intuitively understood in terms of the interaction between the laser fields and atomic medium. As seen in Fig.1a, the Stokes field $E_1$ and anti-Stokes field $E_2$ are produced through two FWM processes, where every Stokes (anti-Stokes) photon generation is obtained by absorbing two coupling (probe) photons and emitting one probe (coupling) photon. In fact, the generated Stokes (or anti-Stokes) field can be equivalently regarded as the result of frequency down- (up-) conversion process through mixing the scattering field with the atomic spin wave S prebuilt by the strong coupling and probe fields (as shown in Fig. 1b). Since the generation of a Stokes (or an anti-Stokes) photon is accompanied with the generation (or annihilation) of an atomic spin-wave excitation, the up-converted frequency component (i.e., anti-Stokes field) is quantum anti-correlated with the down-converted frequency component (i.e., Stokes field), therefore strong bipartite entanglement can be established.

The above idea for producing bipartite entanglement via atomic spin wave can be easily extended to generate multipartite entanglements with any desired order when more applied fields, tuned to the vicinity of the transitions $|1\rangle$-$|3\rangle$ and/or $|2\rangle$-$|3\rangle$, mix with the induced atomic spin wave. For example, when N external fields (including the coupling and probe fields) are applied, 2N-2 entangled fields can be obtained. We demonstrate this concept by realizing tripartite entanglements through scattering an additional mixing field $E_m$ off the atomic spin wave, as shown in Fig. 1. We consider the case of generating three fields $E_1$, $E_2$, and $E_3$ through scattering the coupling, probe, and mixing fields off the atomic spin wave, where the effective Hamiltonian of the system in the interaction picture has the form [15, 16]

$$H_I = \hbar \left[ k_1 \left( a_1^+ S^+ + a_1 S \right) + k_2 \left( a_2^+ S + a_2 S^+ \right) + k_3 \left( a_3^+ S^+ + a_3 S \right) \right].$$

Note that demonstration of tripartite entanglements between fields $E_1$, $E_2$, and $E_4$ can be carried out in the same way. We now define the exchange constant of motion ($C_2$) as

$$C_2 = k_1 \left( a_1^+ S^+ + a_1 S \right) + k_2 \left( a_2^+ S + a_2 S^+ \right) + k_3 \left( a_3^+ S^+ + a_3 S \right).$$

We set $\beta = \sqrt{C_2^2 - \left( k_1^2 + k_3^2 - k_2^2 \right)}$ and obtain the solutions for the operators as follows:

$$S(t) = e^{iC_2 t} \left[ \left( \cos(\beta t) + \frac{iC_2}{\beta} \sin(\beta t) \right) S(0) - \frac{ik_1}{\beta} \sin(\beta t) a_1^+(0) - \frac{ik_2}{\beta} \sin(\beta t) a_2(0) - \frac{ik_3}{\beta} \sin(\beta t) a_3^+(0) \right],$$

$$a_1(t) = -\frac{ik_1}{\beta} \sin(\beta t) e^{-iC_2 t} S^+(0) + \frac{-k_1^2 \beta + k_1^2 \beta \cos(\beta t) e^{-iC_2 t} + ik_1^2 C_2 \sin(\beta t) e^{-iC_2 t} + \beta \left( C_2^2 - \beta^2 \right)}{\beta \left( C_2^2 - \beta^2 \right)} a_1(0)$$

$$+ \frac{-k_1 k_2 \beta + k_1 k_2 \beta \cos(\beta t) e^{-iC_2 t} + ik_1 k_2 C_2 \sin(\beta t) e^{-iC_2 t}}{\beta \left( C_2^2 - \beta^2 \right)} a_2^+(0) + \frac{-k_1 k_3 \beta + k_1 k_3 \beta \cos(\beta t) e^{-iC_2 t} + ik_1 k_3 C_2 \sin(\beta t) e^{-iC_2 t}}{\beta \left( C_2^2 - \beta^2 \right)} a_3(0),$$

$$a_2(t) = -\frac{ik_2}{\beta} \sin(\beta t) e^{iC_2 t} S(0) + \frac{k_1 k_2 \beta - k_1 k_2 \beta \cos(\beta t) e^{iC_2 t} + ik_1 k_2 C_2 \sin(\beta t) e^{iC_2 t}}{\beta \left( C_2^2 - \beta^2 \right)} a_1^+(0)$$

$$+ \frac{k_2^2 \beta - k_2^2 \beta \cos(\beta t) e^{iC_2 t} + ik_2^2 C_2 \sin(\beta t) e^{iC_2 t} + \beta \left( C_2^2 - \beta^2 \right)}{\beta \left( C_2^2 - \beta^2 \right)} a_2(0) + \frac{k_2 k_3 \beta - k_2 k_3 \beta \cos(\beta t) e^{iC_2 t} + ik_2 k_3 C_2 \sin(\beta t) e^{iC_2 t}}{\beta \left( C_2^2 - \beta^2 \right)} a_3^+(0),$$

$$a_3(t) = -\frac{ik_3}{\beta} \sin(\beta t) e^{-iC_2 t} S^+(0) + \frac{-k_1 k_3 \beta + k_1 k_3 \beta \cos(\beta t) e^{-iC_2 t} + ik_1 k_3 C_2 \sin(\beta t) e^{-iC_2 t}}{\beta \left( C_2^2 - \beta^2 \right)} a_1(0)$$

$$+ \frac{-k_2 k_3 \beta + k_2 k_3 \beta \cos(\beta t) e^{-iC_2 t} + ik_2 k_3 C_2 \sin(\beta t) e^{-iC_2 t}}{\beta \left( C_2^2 - \beta^2 \right)} a_2^+(0) + \frac{-k_3^2 \beta + k_3^2 \beta \cos(\beta t) e^{-iC_2 t} + ik_3^2 C_2 \sin(\beta t) e^{-iC_2 t} + \beta \left( C_2^2 - \beta^2 \right)}{\beta \left( C_2^2 - \beta^2 \right)} a_3(0).$$

In this case, the tripartite entanglement of the generated fields $E_1$, $E_2$, and $E_3$ can be demonstrated according to the criterion proposed by van Lock-Furusawa (VLF) [5] with inequalities:

$$V_{12}=V(x_1+x_2)+V(p_1-p_2+g_3p_3)<4,$$

$$V_{13}=V(x_1-x_3)+V(p_1+g_2p_2+p_3)<4,$$

$$V_{23}=V(x_2+x_3)+V(g_1p_1+p_2-p_3)<4,$$

where $V(A)=\langle A^2\rangle-\langle A\rangle^2$ and $g_i$ is an arbitrary real number. Following Ref. [7], we set $g_1=\frac{-(\langle p_1p_2\rangle-\langle p_1p_3\rangle)}{\langle p_1^2\rangle}$, $g_2=\frac{-(\langle p_1p_2\rangle+\langle p_2p_3\rangle)}{\langle p_2^2\rangle}$, and $g_3=\frac{-(\langle p_1p_3\rangle-\langle p_2p_3\rangle)}{\langle p_3^2\rangle}$. Satisfying any pair of these three inequalities is sufficient to demonstrate the creation of tripartite entanglements [5].

Figures 3a-3c depict the evolutions of the VLF correlations as a function of interaction time with $k_2=k_1=1$ and $C_2=30k_1$ under different $k_3$ values. $V_{12}$ is not more than 4 over the whole interaction time for different $k_3$ values, whereas $V_{13}$ and $V_{23}$ display different behaviors when $k_3$ is varied, though they all exhibit periodic oscillations with respect to the interaction time. When $k_1=k_2=k_3=1$ (see Fig. 3b), there exists a wide range of interaction time within which the three inequalities for $V_{12}$, $V_{13}$, and $V_{23}$ are satisfied, indicating that three fields $E_1$, $E_2$, and $E_3$ are CV entangled with each other. Increasing or decreasing $k_3$ would weaken the degree of tripartite entanglements (see Figs. 3a and 3c). Small $k_3$ would lead to the disappearance of the entanglement between $E_3$ and $E_1$ (or $E_2$) fields. By demonstrating the tripartite entanglements between $E_1$, $E_2$, and $E_4$ fields in the same way, one can conclude that all four generated fields ($E_1$, $E_2$, $E_3$, and $E_4$) are entangled with each other.

It should be noted that the current proposed EIT-based scheme for generating multipartite CV entanglements has several distinct features. First, compared to the routinely-employed method to produce multi-field entanglements by using polarizing beam splitters, where the entangled multi-fields are degenerate and suffer from short correlation time (~*ps*), the present configuration can be utilized to generate nondegenerate multiple entangled narrow-band CV fields with long correlation time (~*ms* or even ~*s*), which could be quite useful for quantum memory [10] and quantum networks [3]. Second, as discussed in Ref. [13], in this EIT-based configuration, the atomic spin wave, which mixes with the scattering fields to generate multipartite entanglements, plays a role similar to the polarizing beam splitter used in the traditional way; however, there is a critical difference, that is, the polarizing beam splitter is a linear element, and the atomic medium acts as a nonlinear element for the FWM. Third, by using the experimental setup with a square-box pattern for the laser beams and with different beam polarizations [13], the produced FWM signals can be spatially separated conveniently. Moreover, this configuration with laser beams propagating through the atomic medium in the same direction with small angles among them can benefit from the cancellation of first-order Doppler broadening in the multi-Λ-type system. Finally, the demonstration of tripartite entanglements in the configuration shown in Fig. 1 has provided a clear evidence that more multipartite entangled CV fields can be created via atomic spin wave by applying more scattering fields. The scalability to generate arbitrary number of multipartite entangled fields is the key feature of this proposed scheme.

In conclusion, we have proposed a convenient and flexible way to produce multicolor multipartite CV quantum correlations/anti-correlations and entanglements via atomic spin wave formed by EIT configuration in a multiple-Λ-type atomic system. This method provides a proof-of-principle demonstration of efficiently generating nondegenerate entangled narrow-band multiple fields to any desired order with long correlation time in an atomic ensemble, which may find potential applications in quantum information processing and quantum networks.

## ACKNOWLEDGEMENTS

This work is supported by NBRPC (Nos. 2012CB921804 and 2011CBA00205), the National Natural Science Foundation of China (Nos. 10974132, 50932003, and 11021403), Innovation Program of Shanghai Municipal Education Commission (No. 10YZ10), and Shanghai Leading Academic Discipline Project (No. S30105). Yang's e-mail is yangxih@yahoo.com.cn; M. Xiao's e-mail is mxiao@uark.edu.

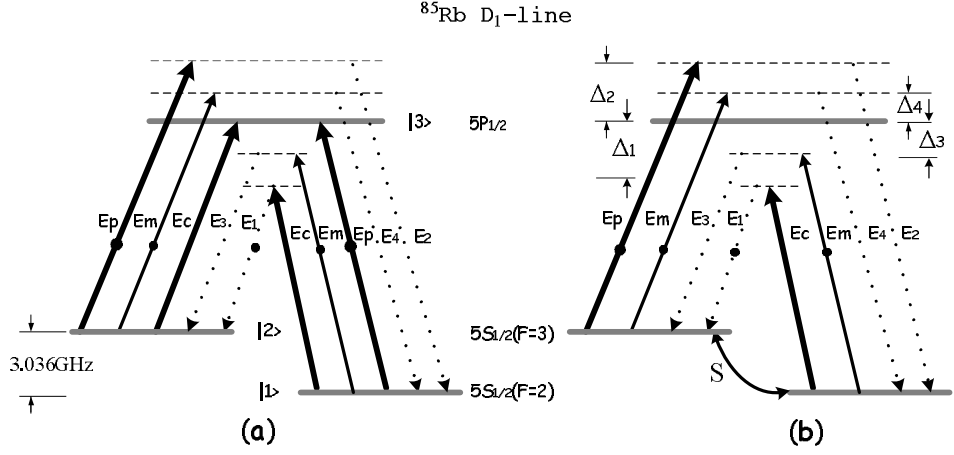

FIG. 1. (**a**) The quintuple-Λ-type system of the $D_1$ transitions in $^{85}$Rb atom coupled by the coupling ($E_c$), probe ($E_p$), and mixing ($E_m$) fields based on the experimental configuration used in Ref. [13], where $E_p$, $E_c$, and $E_m$ fields all drive both $|1\rangle - |3\rangle$ and $|2\rangle - |3\rangle$ transitions, and the corresponding Stokes fields ($E_1$ and $E_3$), and anti-Stokes fields ($E_2$ and $E_4$) are generated through four FWM processes. (**b**) The equivalent configuration of (**a**) with the two lower states driven by the atomic spin wave S induced by the strong on-resonant $E_c$ and $E_p$ fields in the Λ-type EIT configuration.

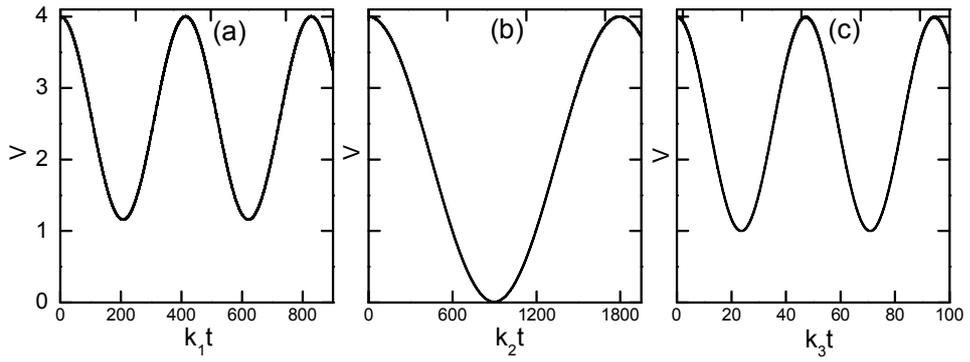

FIG. 2. The evolutions of V as a function of interaction time in terms of the normalized time $k_1 t$ under different $k_2$ values with $k_1=1$ and $C_1=30k_1$. **(a)** $k_2=0.3$, **(b)** $k_2=1.1$, and **(c)** $k_2=3$.

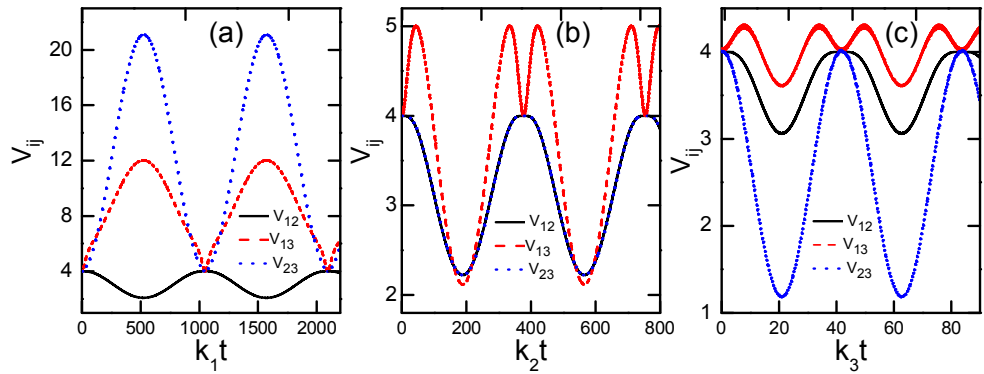

FIG. 3. The evolutions of the VLF correlations $V_{12}$, $V_{13}$, and $V_{23}$ as a function of interaction time under different $k_3$ values with $k_2=k_1=1$ and $C_2=30\ k_1$. **(a)** $k_3=0.6$, **(b)** $k_3=1$, and **(c)** $k_3=3$.